\documentclass[3p,times,procedia]{elsarticle}
\usepackage{nupha_ecrc}

\volume{00}

\firstpage{1}

\journalname{Nuclear Physics A}

\runauth{Roy A. Lacey}

\jid{nupha}

\jnltitlelogo{Nuclear Physics A}

\usepackage{amssymb}

\usepackage{graphicx}
\usepackage{subcaption}
\usepackage[figuresright]{rotating}

\newcommand{\sqsn}{$\sqrt{s_{_{NN}}}$}

\begin{document}

\begin{frontmatter}



\dochead{}

\title{Indications for a critical point in the phase diagram for hot and dense nuclear matter}


\author{Roy A. Lacey}

\address{Depts. of Chemistry \& Physics, Stony Brook University, NY 11794}

\begin{abstract}
Two-pion interferometry measurements are studied for a broad range of collision centralities 
in Au+Au ($\sqrt{s_{NN}}= 7.7 - 200$ GeV)  and Pb+Pb ($\sqrt{s_{NN}}= 2.76$ TeV) collisions.
They indicate non-monotonic excitation functions for the Gaussian emission source radii 
difference ($R^2_{\mathrm{out}} - R^2_{\mathrm{side}}$), suggestive of  reaction trajectories which 
spend a fair amount of time near a  “soft point’’ in the equation of state (EOS) that coincides with the 
critical end point (CEP). A Finite-Size Scaling (FSS) analysis of these excitation functions,  provides
further validation tests for the CEP. It  also indicates a second order phase 
transition at the CEP, and  the values  $T^{\mathrm{cep}} \sim 165$~MeV and $\mu_B^{\mathrm{cep}} \sim 95$~MeV 
 for its location in the ($T, \mu_B$)-plane of the phase diagram. The static critical 
exponents ($\nu \approx 0.66$ and $\gamma \approx 1.2$) extracted via the same FSS analysis, 
place this CEP in the 3D Ising model (static) universality class.  A Dynamic Finite-Size Scaling analysis of the 
excitation functions, gives the estimate $z \sim 0.87$ for the dynamic critical exponent, suggesting that the 
associated critical expansion dynamics is dominated by the hydrodynamic sound mode.   
\end{abstract}


\end{frontmatter}


\section{Introduction}
\label{}

A major goal of the worldwide program in relativistic heavy ion 
research, is to chart the phase diagram for nuclear matter \cite{Itoh:1970,Shuryak:1983zb,Asakawa:1989bq,Stephanov:1998dy}.
Pinpointing the location of the phase boundaries and the critical end point (CEP), in the plane of temperature ($T$) versus 
baryon chemical potential ($\mu_B$), is key to this mapping. Full characterization of  the CEP not only requires 
its location, but also the static and dynamic critical exponents which classify its critical dynamics and thermodynamics, and 
the order of the associated phase transition.

Current theoretical guidance indicates that the CEP belongs to the 3D-Ising [or Z(2)] static universality class 
with the associated critical exponents $\nu \approx 0.63$ and $\gamma \approx 1.2$ \cite{Stephanov:1998dy,exponents}. 
However, the predictions for its location span a broad  swath of the  ($T, \mu_B$)-plane, and do not provide 
a consensus on its location \cite{Stephanov:1998dy}. A recent study which takes account of the non-linear couplings of the 
conserved densities \cite{Minami:2011un} suggests that the CEP's critical dynamics may
be controlled by three distinct slow modes, each characterized by a different value of the dynamic critical 
exponent $z$; a thermal mode ($z_T \sim 3$), viscous mode  ($z_v \sim 2$) and 
a sound mode ($z_s \sim -0.8$).  The phenomena of critical slowing down results from $z > 0$.
The predicted negative value for $z_s$ could have profound implications 
for the CEP search since it implies critical speeding-up for critical reaction dynamics involving {\em only} the 
sound mode. The present-day theoretical challenges emphasize the need for detailed experimental validation 
and characterization of the CEP.
\section{Anatomy of the search strategy for the CEP}
\label{}
The critical point is characterized by several (power law) divergences linked to the divergence of 
the correlation length $\xi \propto \left| t \right|^{-\nu} \equiv \left| T - T^{\mathrm{cep}} \right|^{-\nu}$. 
Notable examples are the baryon number fluctuations 
$\left< (\delta n) \right> \sim \xi^{\gamma/\nu}$, the isobaric heat capacity $C_p \sim \xi^{\gamma/\nu}$ and 
the isothermal compressibility $\kappa_T \sim \xi^{\gamma/\nu}$.
Such divergences suggest that reaction trajectories which are close to the CEP, 
could drive anomalies in the reaction dynamics to give distinct non-monotonic patterns 
for the related experimental observables.
Thus, a current experimental strategy is to carry out beam energy scans which enable a search  
for non-monotonic excitation functions over a broad  domain of  the ($T, \mu_B$)-plane. 
In this work we use the non-monotonic excitation functions for HBT radii combinations that are sensitive to the 
divergence of the compressibility~\cite{Lacey:2014wqa}. 

The expansion of the pion emission source produced in heavy ion collisions, is driven by the sound speed 
$c_s^2 = 1/\rho\kappa_s$, where $\rho$ is the density, $\kappa_s =  \varsigma\kappa_T$ is the isentropic compressibility
and $\varsigma = C_v/C_p$ is the ratio of the isochoric and  isobaric heat capacities. 
%
%
\begin{figure}
\centering
\begin{minipage}{.57\textwidth}
  \centering
  \includegraphics[width=1.0\linewidth]{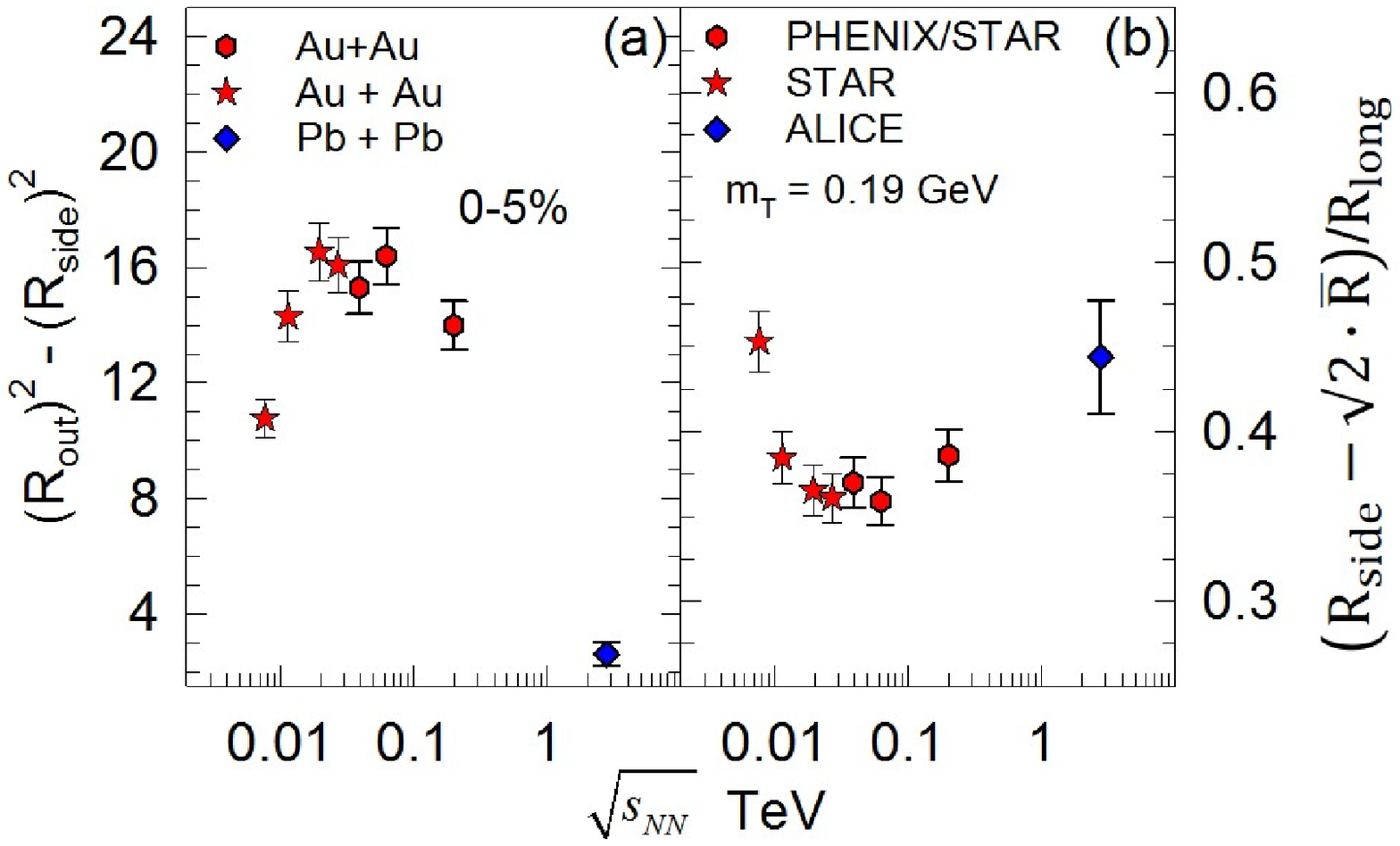}
  \captionof{figure}{(Color online)  The \sqsn\ dependence of (a) $ (R_{\mathrm{out}}^2 - R_{\mathrm{side}}^2)$,\\ 
and  (b) [($R_{\mathrm{side}}- \sqrt{2}\bar{R}$)/$R_{\mathrm{long}}$] \cite{Lacey:2014rxa}.
}
  \label{fig1}
\end{minipage}
\begin{minipage}{.38\textwidth}
  \centering
  \includegraphics[width=1.0\linewidth]{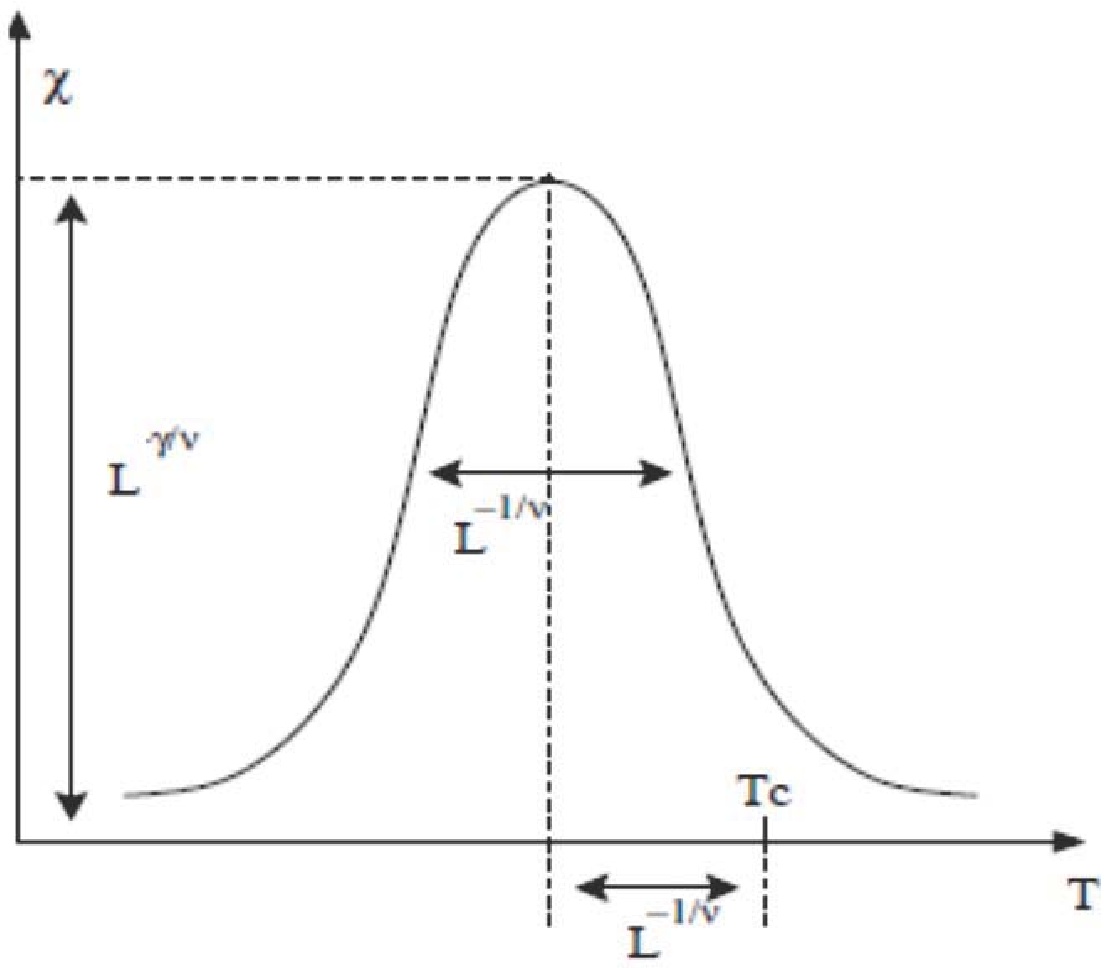}
  \captionof{figure}{Illustration of the Finite-Size ($L$) dependence of the peak position, width and 
magnitude of the susceptibility $\chi$  (see text).
}
  \label{fig2}
\end{minipage}
\end{figure}
%
%
Thus, an emitting source produced in the vicinity of the CEP, would be subject to a precipitous  drop in the 
sound speed and the collateral increase in the emission duration~\cite{Rischke:1996em}, which results from the 
divergence of the compressibility. The space-time information associated with these effects, are encoded
in the Gaussian HBT radii which serve to characterize the emission source.  That is, $R_{{\rm long}}$ is related to the
source  lifetime $\tau$, $(R_{\mathrm{out}}^2 - R_{\mathrm{side}}^2)$ is sensitive to its emission duration 
$\Delta \tau$~\cite{Csorgo:1995bi} (an intensive quantity) and   [($R_{\mathrm{side}}- \sqrt{2}\bar{R}$)/$R_{\mathrm{long}}$] 
gives an estimate for its expansion speed (for small values of $m_T$), where  
$\bar{R}$ is an estimate of the initial transverse size, obtained via Monte Carlo Glauber model 
calculations~\cite{Lacey:2014rxa,Lacey:2014wqa}. Therefore, characteristic convex and concave shapes 
are to be expected for the non-monotonic excitation functions for 
$(R_{\mathrm{out}}^2 - R_{\mathrm{side}}^2)$  and [($R_{\mathrm{side}}- \sqrt{2}\bar{R}$)/$R_{\mathrm{long}}$]
respectively.

These predicted patterns are validated in Figs.~\ref{fig1}(a) and (b). They reenforce the  connection 
between ${(R^2_{\mathrm{out}} - R^2_{\mathrm{side}})}$ and  the compressibility and suggest that 
reaction trajectories spend a fair amount of time near a  “soft point’’ in the EOS that coincides with the CEP.  
We associate  ${(R^2_{\mathrm{out}} - R^2_{\mathrm{side}})}$  
with the susceptibility $\kappa$ and employ Finite-Size Scaling (FSS) for further validation tests, as well as to 
extract estimates for the location of the CEP and the critical exponents which characterize its static and dynamic 
properties.

\begin{figure*}[t]
\includegraphics[width=0.96\linewidth]{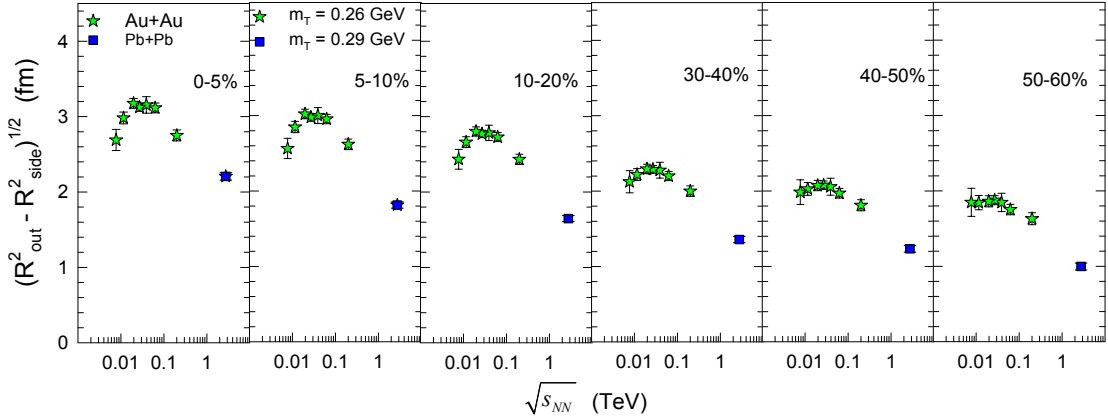}
\vspace{-0.2cm}
\caption{(Color online) ${(R^2_{\mathrm{out}} - R^2_{\mathrm{side}})^{1/2}}$ vs. $\sqrt{s_{NN}}$  
for 0-5\%,  5-10\%, 10-20\%, 30-40\%, 40-50\% and 50-60\% Au+Au  and Pb+Pb collisions
for $m_T = 0.26$~GeV and 0.29~GeV respectively~\cite{Lacey:2014wqa}.
}
\label{fig3}
\vspace{-4pt}
\end{figure*}

\section{Characterization of the CEP via Finite-Size Scaling}
\label{}
For infinite volume systems, $\xi$ diverges near $T^{\mathrm{cep}}$.  Since $\xi \le L$ for a
system of finite size $L^d$ ($d$ is the dimension), only a pseudo-critical point, shifted from the genuine CEP, 
is observed. This leads to a characteristic set of Finite-Size Scaling (FSS) relations
for the magnitude ($\chi ^{\mathrm{max}}_{T}$), width ($\delta T$) and peak position ($t_T$) of 
the susceptibility \cite{Lacey:2014wqa} as illustrated in Fig.~\ref{fig2}; 
$\chi ^{\mathrm{max}}_T(V)  \sim L^{\gamma /\nu}$, $\delta T (V)  \sim  L^{-{1}/{\nu}}$ and 
$t_T(V) \sim  T^{\mathrm{cep}}(V) - T^{\mathrm{cep}}(\infty)   \sim  L^{- {1}/{\nu}}$.
It also leads to the scaling function  $\chi (T, L) = L^{\gamma /\nu}P_{\chi}(tL^{{1}/{\nu}})$,
which results in data collapse onto a single curve.

These scaling relations indicate that even a flawless measurement  can not give 
the precise location of the CEP if it is subject to Finite-Size Effects (FSE) (a crucial 
point which is often missed or ignored). However, they point to specific identifiable dependencies on size (L)  
which can be  leveraged via FSS, to estimate the location of the CEP and its associated 
critical exponents \cite{Lacey:2014wqa}.

Such dependencies can be observed in Fig.~\ref{fig3} where a representative set of  excitation functions,
obtained for the broad selection of centrality cuts, are shown. They  indicate that 
 (i) the magnitude of the peaks decrease with increasing centrality (\%) or decreasing transverse size, 
(ii) the positions of the peaks shift to lower values of $\sqrt{s_{NN}}$ with an increase in centrality and 
(iii) the width of the distributions grow with centrality. 
A Guassian fit was used to extract the peak positions, and widths of the excitation functions, for different system sizes 
characterized by the centrality selections indicated in Fig.~\ref{fig3}; the magnitude 
of ${(R^2_{\mathrm{out}} - R^2_{\mathrm{side}})}$ was evaluated at the extracted peak 
positions as well.  A subsequent FSS analysis (with $L = \bar{R}$), of these peak positions, widths 
and magnitudes was used to obtain estimates for the critical exponents $\nu$ and $\gamma$ and the 
infinite volume $\sqrt{s_{NN}}(\infty)$ value where the de-confinement phase transition first occurs;
${(R^2_{\mathrm{out}} - R^2_{\mathrm{side}})}^{\mathrm{max}} \propto \bar{R}^{\gamma /\nu}$,
and  $\sqrt{s_{NN}}(V)  =  \sqrt{s_{NN}}(\infty)  - k\times \bar{R}^{- {1}/{\nu}}$.
$k$ is a constant and $\delta s \equiv  (\sqrt{s} - \sqrt{s^{\mathrm{cep}}})/\sqrt{s^{\mathrm{cep}}}$
gives a measure of the ``distance"' to the CEP. 

\begin{figure}[t]
\centering
\begin{minipage}{.47\textwidth}
  \centering
	\vspace{6pt}
  \includegraphics[width=0.95\linewidth]{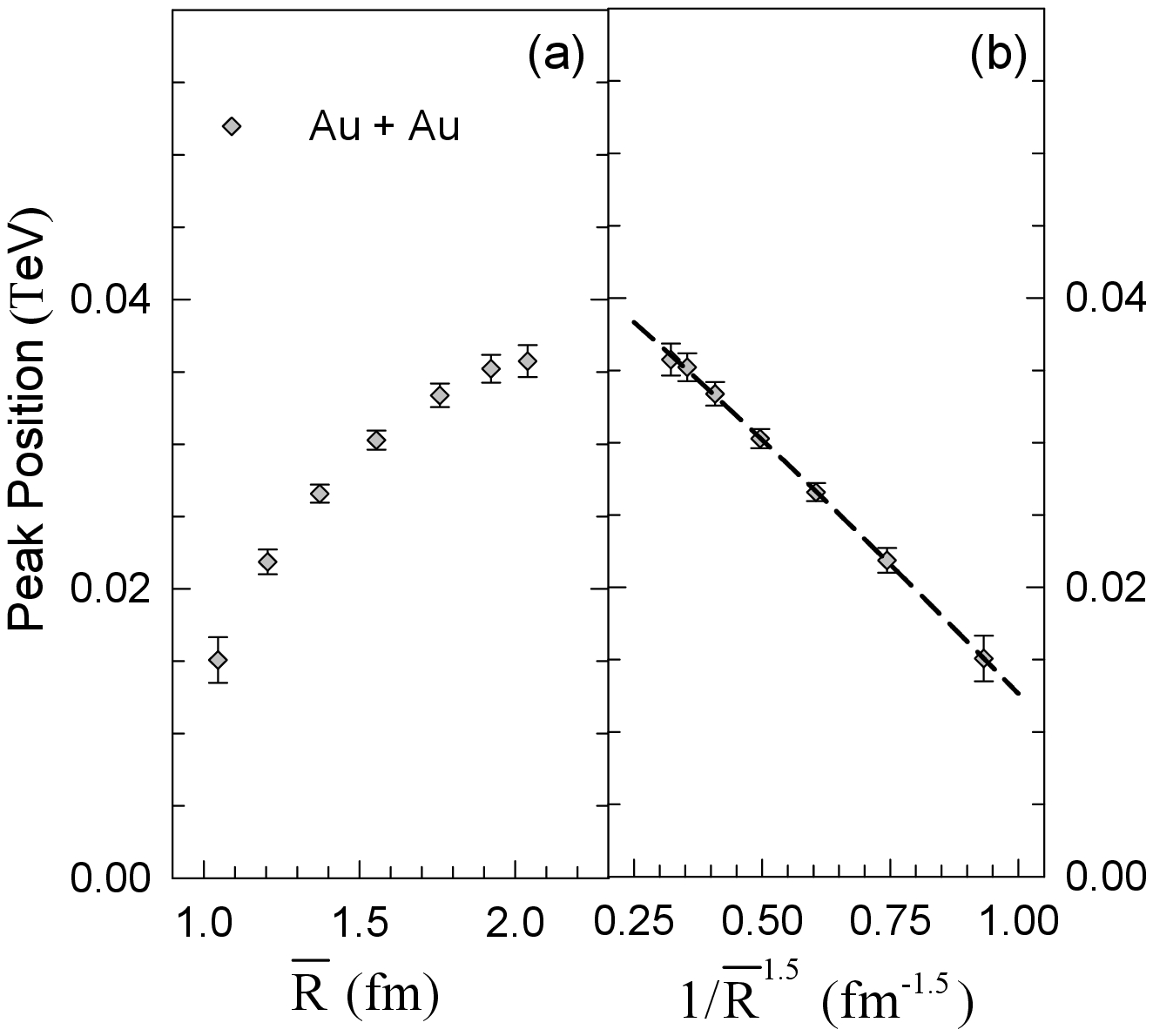}
	\vspace{-4pt}
  \captionof{figure}{(Color online) (a) Peak position vs. $\bar{R}$. (b) Peak position vs. $1/{\bar{R}}^{1.5}$.
The dashed curve in (b) shows the fit to the data in (a).
}
  \label{fig4}
\end{minipage}
\begin{minipage}{.47\textwidth}
  \centering
	\vspace{6pt}
  \includegraphics[width=0.90\linewidth]{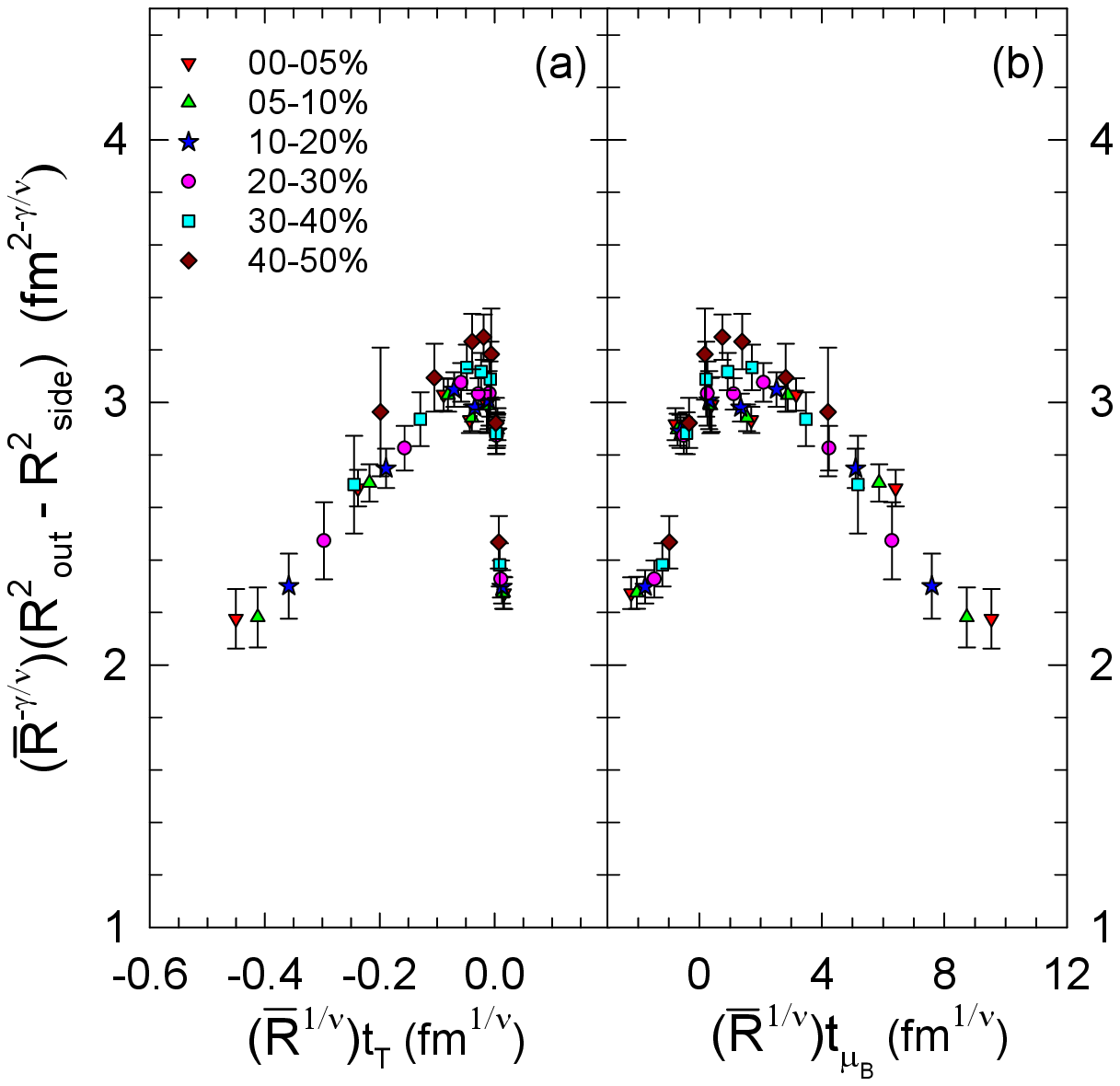}
	\vspace{-4pt}
  \captionof{figure}{(Color online) (a) $\bar{R}^{-\gamma/\nu} (R^2_{\mathrm{out}} - R^2_{\mathrm{side}})$ vs. $\bar{R}^{1/\nu} t_T$.\\
(b)  $\bar{R}^{-\gamma/\nu} (R^2_{\mathrm{out}} - R^2_{\mathrm{side}})$ vs. $ \bar{R}^{1/\nu} t_{\mu_B}$.
}
  \label{fig5}
\end{minipage}
\end{figure}

%
Figure~\ref{fig4} illustrates the FSS test made for the extracted peak positions ($\sqrt{s_{NN}}(V)$). 
The dashed curve in (b), which represents a fit to the data in (a), confirms the expected inverse power law dependence 
of these peaks on $\bar{R}$. The fit gives the values $\sqrt{s_{NN}}(\infty)  =  47.5 \pm 1.5$~ GeV and $\nu = 0.67 \pm 0.05$.
The same value for $\nu$ was obtained via an analysis of the widths. The estimate $\gamma = 1.15 \pm 0.065$,
was obtained from FSS of the the magnitudes of the excitation functions. 
The extracted values for the critical exponents indicate that the deconfinement phase transition at the CEP is 
second order, and places it in the 3D Ising model (static) universality class. The extracted 
value $\sqrt{s_{NN}}(\infty)  =  47.5$~GeV  was used  in conjunction with 
the parametrization for chemical freeze-out in Ref.~\cite{Cleymans:2006qe}, to obtain the 
estimates $\mu_B^{\mathrm{cep}} \sim 95$~MeV and $T^{\mathrm{cep}} \sim 165$~MeV for its 
location in the ($T, \mu_B$)-plane.
%
%

A crucial crosscheck for the location of the CEP and its associated critical exponents, 
is to use the FSS function to demonstrate data collapse onto a single curve for the extracted 
values of $T^{\mathrm{cep}}$,  $\mu_B^{\mathrm{cep}}$ and the critical exponents $\nu$ and $\gamma$;
$\bar{R}^{-\gamma/\nu} (R^2_{\mathrm{out}} - R^2_{\mathrm{side}})  \mathrm{\: vs.\:}  \bar{R}^{1/\nu} t_T$ and 
$\bar{R}^{-\gamma/\nu} (R^2_{\mathrm{out}} - R^2_{\mathrm{side}})  \mathrm{\: vs.\:}  \bar{R}^{1/\nu} t_{\mu_B}$
%
where $t_T = (T - T^{\mathrm{cep}})/T^{\mathrm{cep}}$ and  $t_{\mu_B} = (\mu_B - \mu_B^{\mathrm{cep}})/\mu_B^{\mathrm{cep}}$
are the reduced temperature and baryon chemical potential respectively.
The efficacy of this crosscheck is illustrated in Fig.~\ref{fig5} where data collapse onto a single curve is indicated 
for the RHIC excitation functions shown in Fig.~\ref{fig3}.
Here, the parametrization for chemical freeze-out \cite{Cleymans:2006qe} was used in conjunction with  $\mu_B^{\mathrm{cep}}$ 
and  $T^{\mathrm{cep}}$ to determine the required $t_T$ and  $t_{\mu_B}$ values from  the $\sqrt{s_{NN}}$ values 
plotted in Fig.~\ref{fig3}. Figs.~\ref{fig5}(a) and (b) also validate the expected  trends for reaction trajectories 
in the ($T, \mu_B$) domain  which encompass the CEP. 
That is, the scaled values of $(R^2_{\mathrm{out}} - R^2_{\mathrm{side}})$ peaks at $t_T \approx 0$ and $t_{\mu_B} \approx 0$, 
and show the collateral fall-off  for $t_{T, \mu_B} < 0$ and $t_{T, \mu_B} > 0$.

%
The susceptibility diverges at the CEP, so relaxation of the order parameter could be anomalously slow, {\em i.e.}
$\tau \sim \xi^z$, where $z$ is the dynamic critical exponent. Such an anomaly can lead to significant attenuation of 
the signals associated with the CEP ($\xi \sim \tau^{1/z}$) for critical dynamics driven by the thermal ($z_T \sim 3$) and 
viscous ($z_v \sim 2$) slow modes, especially for the short-lived processes of interest. This phenomena of critical 
slowing down would switch to critical speeding up for critical dynamics driven by the sound mode ($z_s \sim -0.8$).
 Thus, a rudimentary knowledge of the magnitude and the sign of the dynamic critical exponent/s can significantly 
enhance experimental studies of the CEP. The Dynamic Finite Size scaling function  can be expressed as;
$\chi (L, T, \tau) = L^{\gamma /\nu}f(L^{1/\nu}t_T, \tau L^{-z})$.
For $T \sim T^{\mathrm{cep}}$ it simplifies to the expression $\chi (L, T, \tau) = L^{\gamma /\nu}f(\tau L^{-z})$
which is observed to scale the data for 
$\bar{R}^{-\gamma/\nu} (R^2_{\mathrm{out}} - R^2_{\mathrm{side}})  \mathrm{\: vs.\:}  R_{\mathrm{long}}\bar{R}^{-z}$
and give the estimate $z \sim 0.87$.  Here, $R_{\mathrm{long}}$ is used as a proxy for $\tau$.

%
In summary, we have investigated  the centrality dependent excitation functions for the Gaussian emission 
source radii difference ($R^2_{\mathrm{out}} - R^2_{\mathrm{side}}$), obtained from  two-pion interferometry 
measurements in Au+Au ($\sqrt{s_{NN}}= 7.7 - 200$ GeV)  and Pb+Pb ($\sqrt{s_{NN}}= 2.76$ TeV) 
collisions, to characterize the CEP. The observed centrality dependent  non-monotonic excitation functions, 
validate characteristic finite-size scaling patterns that are consistent with a deconfinement phase transition 
and the critical end point. A Finite-Size Scaling analysis of these data indicates a second order phase transition at 
a CEP located at  $T^{\mathrm{cep}} \sim 165$~MeV and $\mu_B^{\mathrm{cep}} \sim 95$~MeV in the 
($T, \mu_B$)-plane of the phase diagram. 
The critical exponents ($\nu = 0.67 \pm 0.05$ and $\gamma = 1.15 \pm 0.065$)
extracted in the same FSS analysis, places the CEP in the 3D Ising model (static) universality class.  
An initial estimate of $z \sim 0.87$ for the dynamic critical exponent is incompatible with the commonly 
assumed Model H dynamic universality class ($z \sim 3$) assigned to critical expansion dynamics.





\bibliographystyle{elsarticle-num}
\bibliography{CEP_Lacey-QM2015}

\begin{thebibliography}{10}
\bibitem{exponents}
The critical exponents satisfy several equalities. Thus, only two of the four
  exponents are independent.
\end{thebibliography}







\end{document}